\documentclass[pra,twocolumn,showpacs,groupedaddress,nofootinbib]{revtex4}
\usepackage{amsmath,amssymb,amsfonts}
\usepackage{graphicx}
\usepackage{txfonts}

\begin{document}

\title{Magnetic transport apparatus for the production of ultracold atomic gases in the vicinity of a dielectric surface}

\author{S. H\"{a}ndel}
\affiliation{Department of Physics, Durham University, Durham DH1 3LE, United Kingdom}
\author{A. L. Marchant}
\affiliation{Department of Physics, Durham University, Durham DH1 3LE, United Kingdom}
\author{T. P. Wiles}
\affiliation{Department of Physics, Durham University, Durham DH1 3LE, United Kingdom}
\author{S. A. Hopkins}
\affiliation{Department of Physics, Durham University, Durham DH1 3LE, United Kingdom}
\author{S. L. Cornish}
\affiliation{Department of Physics, Durham University, Durham DH1 3LE, United Kingdom}

\date{\today}

\begin{abstract}
We present an apparatus designed for studies of atom-surface interactions using quantum degenerate gases of $^{85}$Rb and $^{87}$Rb in the vicinity of a room temperature dielectric surface. The surface to be investigated is a super-polished face of a glass Dove prism mounted in a glass cell under ultra-high vacuum (UHV). To maintain excellent optical access to the region surrounding the surface magnetic transport is used to deliver ultracold atoms from a separate vacuum chamber housing the magneto-optical trap (MOT). We present a detailed description of the vacuum apparatus highlighting the novel design features; a low profile MOT chamber and the inclusion of an obstacle in the transport path. We report the characterization and optimization of the magnetic transport around the obstacle, achieving transport efficiencies of 70\% with negligible heating. Finally we demonstrate the loading of a hybrid optical-magnetic trap with $^{87}$Rb and the creation of Bose-Einstein condensates via forced evaporative cooling close to the dielectric surface.
\end{abstract}


\maketitle

\section{Introduction}

The exquisite control with which ultracold and quantum degenerate atomic gases can be manipulated has enabled the study of a diverse range of physical phenomena~\cite{inguscio_stringari_book_99,inguscio_ketterle_book_07,pethick_smith_08}, in many cases with interesting parallels in other fields of physics~\cite{lewenstein_sanpera_07,buluta_nori_09,saffman_walker_10}. The same exquisite control offers new possibilities in a number of precision measurement applications~\cite{chin_flambaum_09,roati_deMirandes_04,carusotto_pitaevskii_05,gustavsson_haller_08} which often complement the established traditional approaches. A good example is the search for new fundamental physics at short length scales through the observation of non-Newtonian gravitational forces~\cite{Geraci2008}, where several measurement schemes exploiting ultracold atoms have now been proposed~\cite{Sandoghdar1992,Dimopoulos2003,Onofrio2006,Wolf2007}. Indeed, a number of recent experiments have begun to investigate ultracold atomic gases near to room-temperature surfaces~\cite{Hinds1993,Aspect1996,Pasquini2004,Harber2005,mohapatra_unnikrishnan_06,Obrecht2007,Sorrentino2009,Bender2010,Stehle2011} where it is first necessary to fully understand the attractive Casimir-Polder force~\cite{Casimir1948}. However, combining the technology required to produce and manipulate a quantum degenerate gas with a high quality surface presents a number of technical challenges.

In this paper we describe an apparatus with several unique features designed to overcome these challenges in order to study $^{85}$Rb Bose-Einstein condensates with tunable atomic interactions~\cite{Cornish2000,Altin2010} near to a room-temperature surface. The surface to be investigated is a super-polished face of a glass Dove prism mounted in a glass cell under ultra-high vacuum (UHV). The use of a Dove prism permits the creation of an optical evanescent-wave potential in order to reflect or trap atoms near to the surface. The surface is oriented vertically with the aim of studying the quantum reflection~\cite{Shimizu2001,Druzhinina2003,Pasquini2004,Zhao2010} of low-velocity bright matter-wave solitons~\cite{Cornish2009} launched horizontally in an optical waveguide aligned through the prism~\cite{Marchant2011}. To preserve optical access close to the surface magnetic transport~\cite{Greiner2001,Lewandowski2003} is used to deliver ultracold atoms to the UHV glass cell, thereby confining the hardware required to collect atoms in a magneto-optical trap (MOT) to a separate chamber. The MOT chamber has been designed to use the full diameter of a standard CF40 viewport whilst minimizing the vertical profile to facilitate the magnetic transport. The two chambers are connected via a differential pumping tube housing a right angle prism which serves as an obstacle to shield the super-polished surface from stray rubidium following ballistic trajectories whilst also providing additional optical access. The simple application of an appropriate magnetic bias field allows the ultracold atomic gas to be safely transported around the obstacle. Here we present the optimization of this novel transport scheme, together with a detailed description of the key elements of the apparatus. 

The structure of the rest of the paper is as follows. In section~\ref{sec:VacuumSystem} we describe the design and construction of the vacuum system, starting with the motivations behind its design before providing detailed descriptions of its three main parts: the MOT chamber, the differential pumping stage and the UHV glass cell. In section~\ref{sec:Magnetictransportoveranobstacle} we introduce the basic mechanics of the magnetic transport system before describing the specific loading of ultracold atoms from the MOT into the transport trap and their subsequent transport through the differential pumping stage, over an obstacle prism and to the glass cell. We then present the results of calculations and experiments used to analyze and optimize the efficiency of the transport process and establish a range of optimum conditions for efficient transport of atoms. In section~\ref{sec:HybridTrap} we describe use of the complete apparatus to load a hybrid optical-magnetic trap and then to produce a Bose-Einstein condensate of $^{87}$Rb atoms, which is an important milestone on our route to study quantum degenerate gases of rubidium in the vicinity of a surface. Section~\ref{sec:Conclusion} is a summary and outlook. 

\section{Vacuum system}\label{sec:VacuumSystem}
\begin{figure*}
	\centering
		\includegraphics[width=0.75\textwidth]{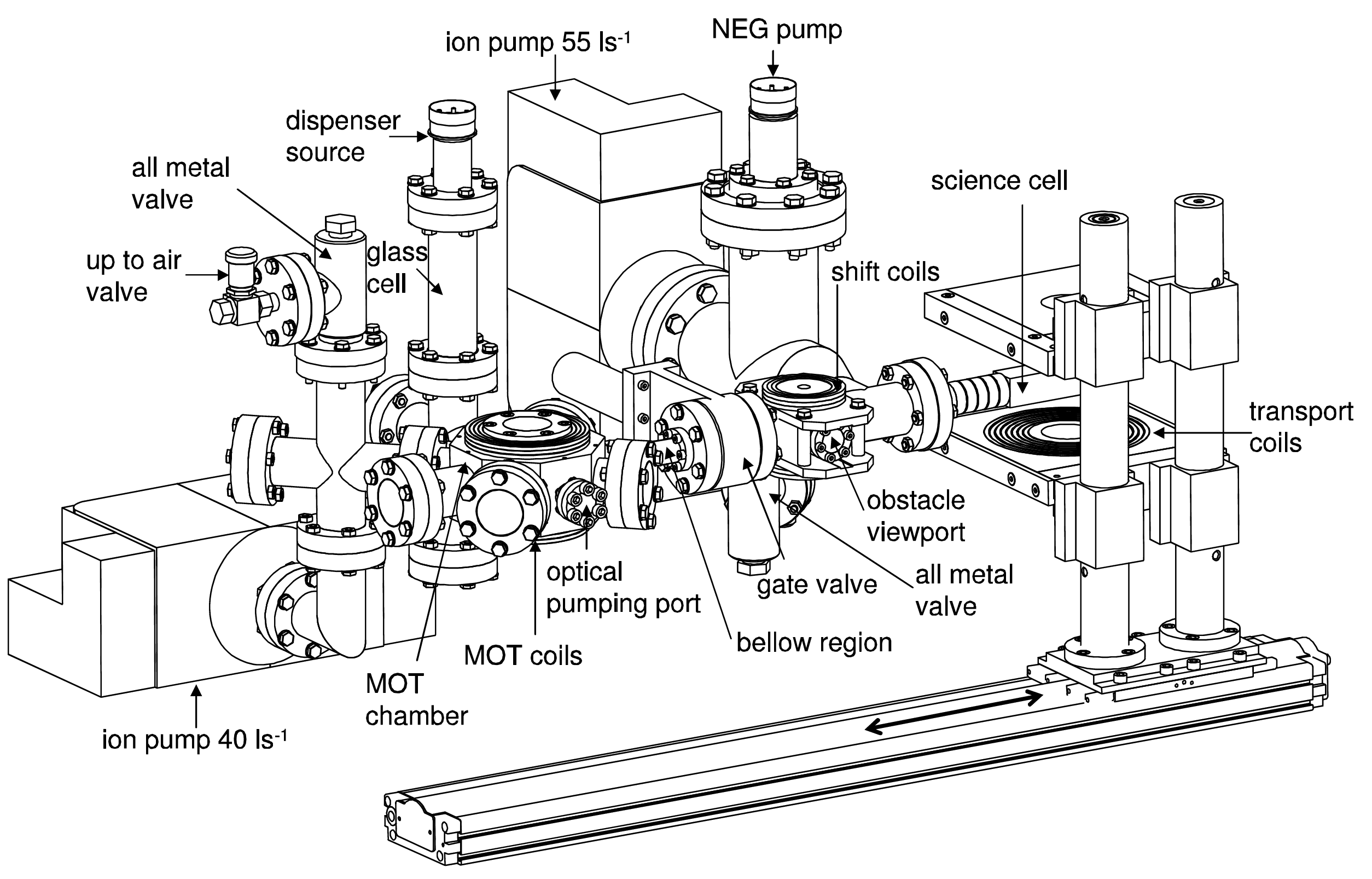}
	\caption{Overview of the complete vacuum system. The octagonal MOT chamber is connected via a differential pumping stage to the glass science cell. The length of the system from the center of the MOT chamber to the center of the glass cell is 51~cm. Atoms are transported within the vacuum system in a magnetic quadrupole trap which is mounted on a motorized translation stage. The line of sight between the MOT chamber and the science cell is blocked by a glass prism inserted into the transport path. In order to transport the atoms over the glass prism, a pair of shift coils is attached to the vacuum system. The glass cell houses a super-polished prism and is surrounded by a coil assembly (not shown for clarity) used for generating a static quadrupole potential and a magnetic bias field.}
	\label{fig:fig1}
\end{figure*}
The primary goal of the apparatus is to enable studies of atom-surface interactions in an ultra-high vacuum (UHV) environment, using a super-polished glass Dove prism as the test surface and ultracold rubidium atoms in variable-geometry optical traps close to this surface. When designing such an apparatus, a first important design factor is to create good optical access to this near-surface test region as it enables flexibility for future optical trap geometries, which may include waveguide beams, crossed dipole trap beams at various angles, optical lattice beams and evanescent wave beams at the prism surface. A second design factor, the desirability of a small vacuum chamber, arises because our planned experiments will involve both magnetic trapping and exploitation of atomic Feshbach resonances. These require the creation at the test region of large magnetic field gradients ($\gtrsim$200~G\,cm$^{-1}$) or bias fields ($\gtrsim$100~G) and this is most easily achieved with small, closely spaced coil pairs placed outside but adjacent to the chamber. A third design factor stems from the need to keep the super-polished prism surface as free as possible from contamination by stray rubidium. Thus it is desirable to avoid any atomic beam directed into the glass cell and prism region as is the case in systems using a Zeeman slower \cite{phillips_metcalf_82} or a double MOT \cite{aminoff_steane_93}. All three design factors mentioned above are satisfied by the choice for the UHV vacuum chamber of a small rectangular glass cell containing the Dove prism surface. To preserve the excellent optical access of such a cell, we use a separate MOT chamber for the initial cooling and trapping of atoms, connected by a narrow differential pumping stage to the glass cell. Thus the optics and coils and other paraphernalia associated with the MOT are displaced and do not block lines of sight around the glass cell.

The delivery of atoms from the MOT chamber to the glass cell is accomplished by a magnetic transport system whereby a movable magnetic quadrupole trap formed by an anti-Helmholtz coil pair is mounted on a motorized translation stage. Only the needed atoms are brought, in a controlled manner, into the glass cell region. Furthermore the use of magnetic transport allows the placement of an obstacle in the direct line of sight path from the MOT chamber to the glass cell, thus preventing stray rubidium atoms from reaching the Dove prism via ballistic trajectories. The moving transport coils enclose, at opposite ends of their travel range, both the MOT chamber and the glass cell apparatus (see fig.~\ref{fig:fig1}). Again considering that magnetic field and gradients scale inversely with the physical size, it is desirable to keep the magnetic transport coils as small as possible and consequently we must also make the MOT chamber as compact as possible. We believe that the design presented here represents the most compact arrangement that can be achieved using the full optical access of a standard CF40 size viewport and will therefore be of great interest to the growing number of groups choosing to adopt the magnetic transport technique.

We now describe in detail the construction of the three main parts of the vacuum system, the MOT chamber, the differential pumping stage and the UHV glass cell in accordance with the above design.

\subsection{MOT chamber}
\begin{figure}
	\centering
		\includegraphics[width=0.48\textwidth]{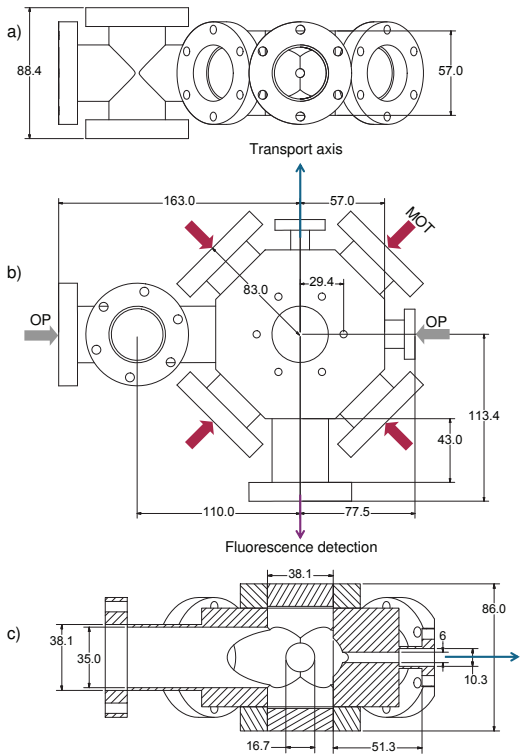}
	\caption{Drawing of the MOT chamber (all dimensions are in mm). a) Side view along the transport axis. The height of the MOT chamber body is 57~mm. The small central hole is a 6~mm diameter exit hole for the transported atoms. b) Bird-eye's view of the octagonal shaped MOT chamber. Four DN40 CF-flanges allow optical access for the horizontal MOT beams. A DN16 CF-flange gives optical access for the optical pumping beam. Another DN40 CF-flange is used for fluorescence detection. c) Cross sectional view of the MOT chamber along the transport axis. The overall height of the system is minimized by using counterbored view ports on the top and bottom of the chamber.}
	\label{fig:fig2}
\end{figure}
The MOT chamber is designed to have the lowest possible height whilst still maintaining access for large diameter laser cooling beams. It consists of a custom octagonal chamber constructed of type 316LN stainless steel, chosen for its low magnetic permeability. The chamber has six large CF40 viewports for laser cooling that have a nominal tube inner diameter of 35.0~mm. The four horizontal viewports are attached via short extruding pipes, but the upper and lower viewports are directly attached to the chamber. There are additional viewports for an optical pumping beam and fluorescence detection and also an outlet port for the atoms (the magnetic transport axis). All viewports have a double-sided anti-reflection coating~\cite{Viewport}, allowing for MOT beams with a 30~mm diameter, i.e. large enough to capture $>10^9$ atoms in the MOT within a few seconds. Fig.~\ref{fig:fig2} presents an engineering drawing of the MOT chamber. In fig.~\ref{fig:fig2}\,(a) a crosscut of the MOT chamber is shown, looking along the axis of magnetic transport [blue arrow in fig.~\ref{fig:fig2}\,(b)]. The overall height of the chamber body is 57~mm, which increases to 86~mm [see fig.~\ref{fig:fig2}\,(c)] when the counterbored top and bottom viewports (Caburn, ZCVP-40-CBORE) are attached. The counterbores are sufficiently deep to entirely absorb the M6 boltheads, such that the 86~mm height represents the full restriction for the inner separation of the coils used for the magnetic transport. The MOT coils are wound in such a way to sit flush on the chamber body surrounding the counterbored viewports. The body holes themselves are 9 mm deep, leaving only 2 mm of material before the inner chamber. In fig.~\ref{fig:fig2}\,(b) a bird-eye's view of the MOT chamber is shown including the MOT beams and the optical pumping (OP) beams. The viewport marked with the purple arrow is used for fluorescence detection to monitor the MOT loading performance. In fig.~\ref{fig:fig2}\,(c) a cross cut of the MOT chamber is shown; the blue arrow marks the direction of the magnetic transport. The transport axis starts from the center of the MOT chamber with a bored 6 mm diameter hole of length 30 mm leading to an externally welded tube with an inner diameter of 10.3~mm, leading in turn to a DN16 flange for connection to the next differential pumping stage. 
The MOT chamber is pumped by an ion getter pump (VacIon plus 40 starcell pump) which is attached via an elbow to a four way cross (see fig.~\ref{fig:fig1}). Attached to this four-way cross is an all metal valve (VG Scienta, ZCR40R) in combination with an up-to air valve (Swagelok, 275-5TV). The MOT chamber also has a second four way cross welded into it, along the optical pumping axis. This cross contains the source of rubidium for the MOT chamber. In each of the top and bottom of the cross six alkali metal dispensers (SAES Getter, NF3.4F12FT10+10) are fixed to the tips of a six pin molybdenum electrical feed through. A glass cell between the top and bottom dispensers may be used to implement the light induced atomic desorption (LIAD) technique~\cite{Klempt2006}. 

\subsection{Differential pumping and obstacle}
A differential pumping stage connects the MOT chamber with the glass cell. It has two functions. Firstly it provides a high vacuum route for conveyance of the cold atoms in the magnetic transport trap from the MOT chamber into the UHV glass cell region. Secondly its low overall  conductance of 0.3~l\,s$^{-1}$ maintains a differential pressure of $>10^2$ between the MOT and glass cell regions, thus allowing swift loading of the MOT simultaneously with a long lifetime for trapped atoms in the glass cell. The entire transport axis has several sections arranged in a straight line: starting from the MOT chamber there is a 6~mm diameter, 30~mm long section connecting to a 10.3~mm diameter, 22 mm long vacuum tube. These two sections end with a DN16 flange at 77.4~mm from the center of the MOT chamber; this is connected to a 76~mm long flexible steel bellows with an inner diameter of 16.7~mm (Caburn, 400000). At the end of the bellows a 5~mm diameter copper aperture with a thickness of 2~mm was inserted. The aperture diameter of 5~mm is just large enough to allow a cloud of magnetically trapped atoms at a temperature of 200~$\mu$K to pass.

Next in line, a non-magnetic gate valve (VAT01032-Ce01-ARF 1) seals the MOT chamber from the glass cell and enables separate pumping of the two halves when commissioning the vacuum system. It can also be used to maintain the high vacuum in either of them, for instance if the MOT chamber vacuum was to be broken for changing the alkali metal dispensers. After the gate valve a wider section of 35 mm diameter steel pipe, 160 mm long, leads to a DN40 flange where a glass-to-metal circular section, 28 mm diameter, 85 mm long finally leads to the square section glass cell.

\begin{figure}
	\centering
		\includegraphics[width=0.48\textwidth]{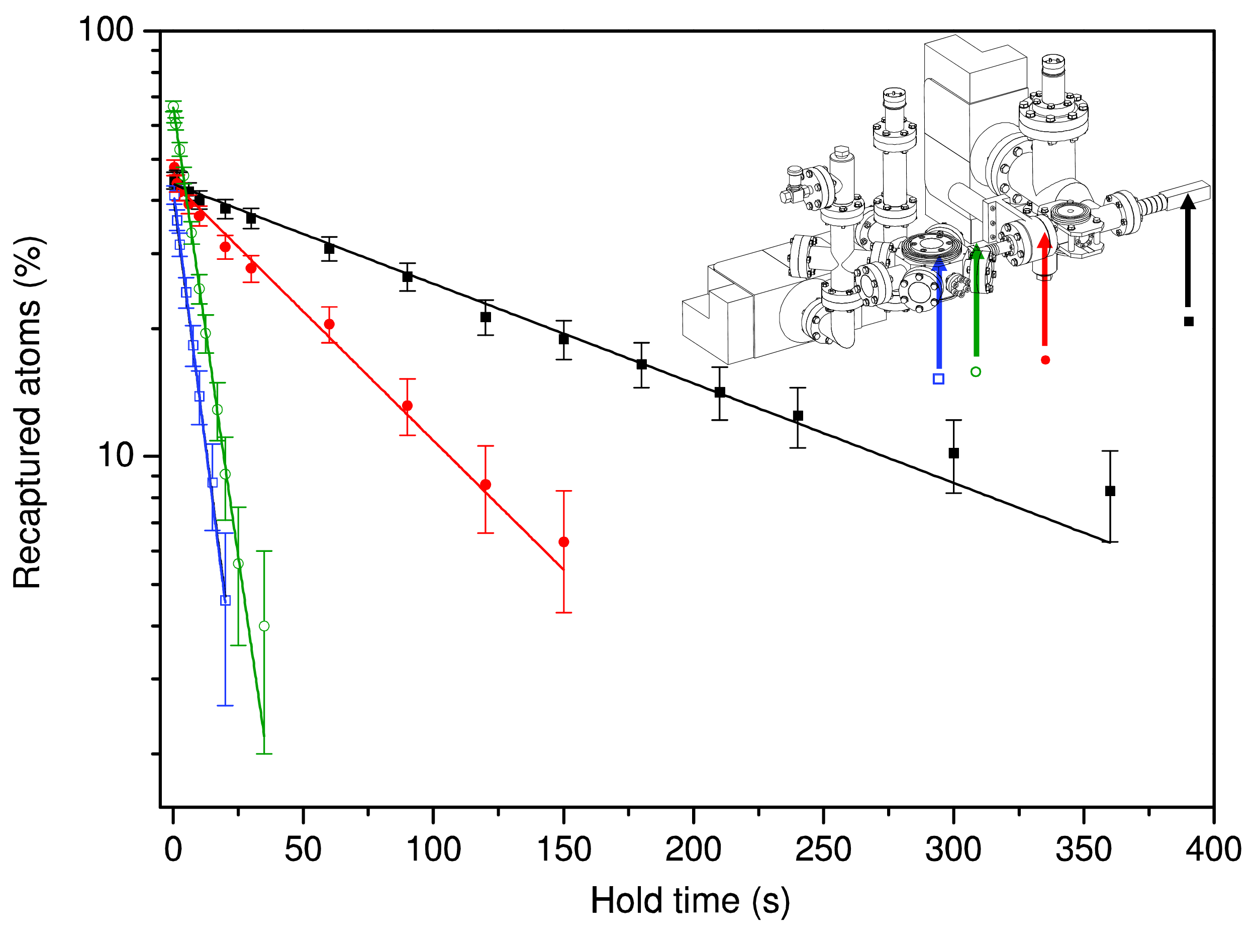}
	\caption{Lifetimes of a cold atomic cloud for four positions along the transport axis of the vacuum system. Atoms were loaded into a $180\,\mbox{G$\,$cm$^{-1}$}$ transport trap and then moved to one of the four positions along the transport axis and held for a variable time. The atomic cloud was then transported back to the initial start position and the number of atoms recaptured in the MOT was recorded using fluorescence detection. Fitting the data with a single exponential decay returned lifetimes of: 10(1)~s (blue, open squares) in the MOT chamber, 9(1)~s (green, open circles) at the start of the 6~mm tube section, 72(5)~s (red, full circles) in the bellows section just before the gate valve and 186(9)~s (black, full squares) in the science cell.}
	\label{fig:fig3}
\end{figure}
In order to characterize the quality of the vacuum, the lifetime of atoms in the magnetic transport trap was measured at various positions along the vacuum system. Fig.~\ref{fig:fig3} shows the results of these measurements. For all these lifetime measurements the Rb dispensers were run at a current of 3.5(1)~A, which corresponds in our system to a MOT loading rate of 1.3(2)$\times 10^8$ atoms\,s$^{-1}$. The first lifetime measurement was taken directly in the center of the MOT chamber. For this purpose atoms were captured from the MOT in a magnetic trap with an axial gradient of 45~G\,cm$^{-1}$ which was then adiabatically stiffened to 180~G\,cm$^{-1}$. The atoms are held for a variable amount of time in this trap and then released and recaptured by the MOT. Fluorescence detection was used to determine the recaptured atom number. The lifetime in the MOT chamber (blue arrow, open square) was limited by collisional losses to 10 seconds due to the relatively high rubidium background pressure.  The second lifetime was measured by moving the magnetic trap to 3.5~cm away from the MOT center, which is at the entrance of the 6~mm hole at the start of the differential-pumping stages. Here the lifetime (green arrow, open circle) was still short (9(1)~s). A third lifetime was taken at a transport distance of 15~cm, i.e. just before the 5~mm aperture next to the gate valve. Here the lifetime increased to 72(5)~s as the trap is approaching the ultra-high vacuum region. Fully transporting the trapped atoms into the science cell (black arrow, full square) resulted in a lifetime of 186(9)~s, which is suitable for RF-induced evaporative cooling.
\begin{figure}
	\centering
		\includegraphics[width=0.48\textwidth]{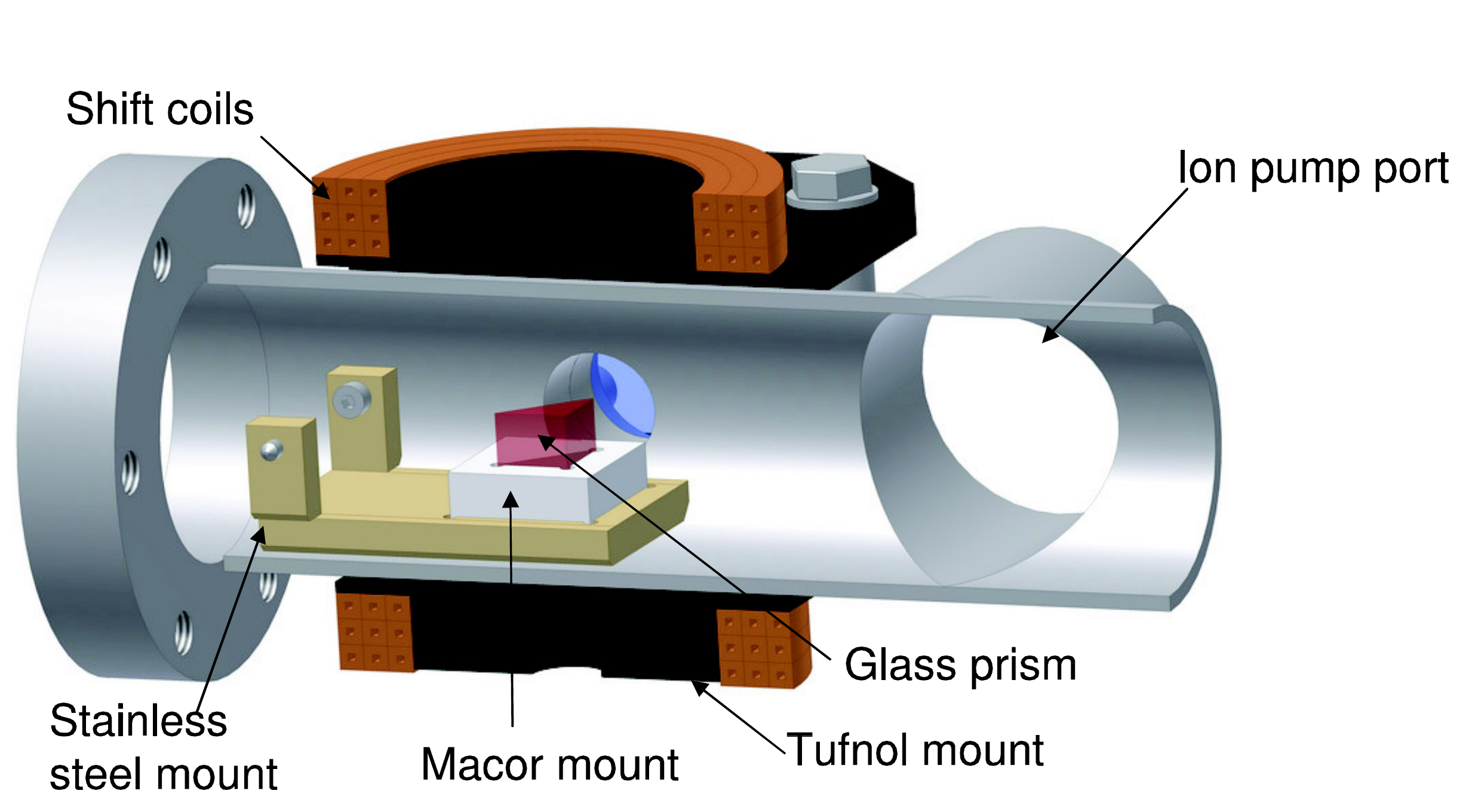}
	\caption{Setup of the obstacle prism in the vacuum system. A glass prism, embedded in a macor mount is mounted into the transport path. The mount is held with two rounded down screws in the steel vacuum tube. The prism blocks the line of sight between MOT chamber and glass cell and allows injection of further beams by total internal reflection at its back face.}
	\label{fig:fig4}
\end{figure}

Along the path of the transport we block the line of sight between the MOT chamber and the science cell with an uncoated glass 45$^{\circ}$ prism (Thorlabs, PS610). This serves two purposes. Firstly it optically separates the two chambers whilst allowing beams to be injected into the obstacle viewport (fig.~\ref{fig:fig1}) along the transport axis towards the science cell, or equally well, the reverse of this. The beams undergo total internal reflection at the rear prism surface and are hence deflected through 90$^{\circ}$. Secondly the obstacle prism blocks any direct flux of rubidium from the MOT chamber to the glass cell, thus protecting the super-polished surface of the Dove prism. In fig.~\ref{fig:fig4}, a schematic setup of the prism in the transport path is shown. The prism is held in a stainless steel mount which is held by two round-headed screws in the steel vacuum tube. 

\subsection{Science chamber}
The science chamber consists of a rectangular glass cell (made by Starna UK) combined with a custom made pumping chamber. The main vacuum pumping in the science chamber is performed by a (55~l\,s$^{-1}$) ion getter pump (VacIon plus 55 starcell pump), which is attached to the science chamber by a DN63CF flange. This is assisted by a non-evaporative getter (CapaciTorr-D 400-2 pump). The rectangular glass cell is made from optically-contacted, 2~mm thick, fused silica and has internal dimensions of 20$\times$20$\times$83~mm. A cylindrical graded index glass-to-metal section increases the overall length of the cell from its flat glass end to the flange to 17~cm. The cell allows excellent optical access, whilst at the same time being compact which facilitates the adjacent fitting of several magnetic field coils. In our experiment the coils in this region comprise the following: quadrupole coils to produce a magnetic trap with gradients up to 400~G\,cm$^{-1}$, bias coils to produce fields up to 600~G, levitation coils to magnetically cancel the effect of gravity on the atoms and shim coils to null the local earth field. 

\begin{figure}
	\centering
		\includegraphics[width=0.48\textwidth]{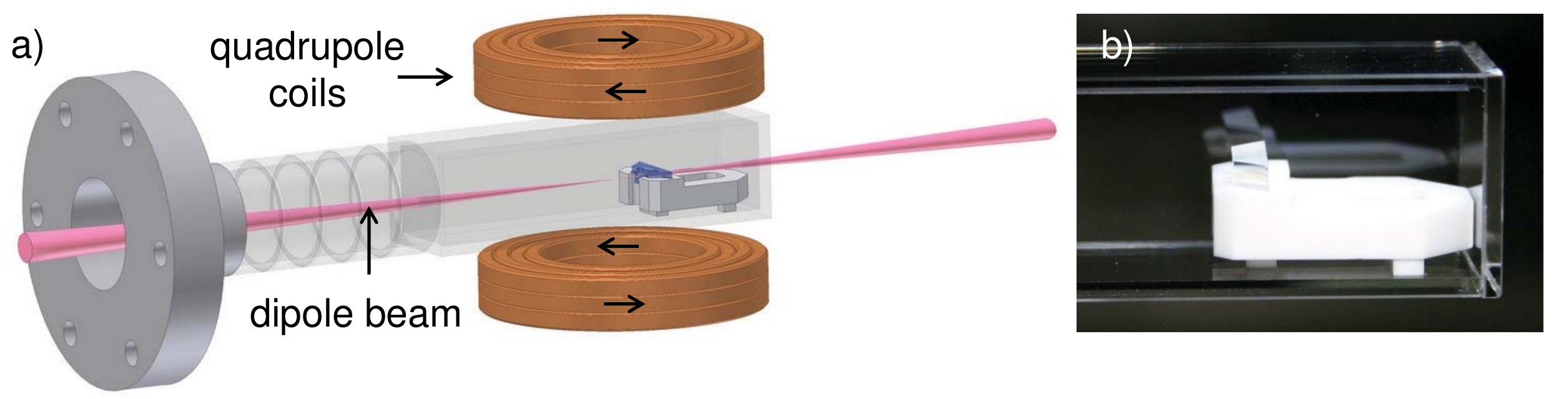}
	\caption{ a) Schematic of the science cell setup showing the glass cell, the Dove prism and a set of quadrupole coils. A dipole trapping beam enters the glass cell through the rear face of the prism and subsequently exits the vacuum chamber via the obstacle viewport. b) Photograph of the Dove prism in the glass cell. The super-polished glass prism sits in a ceramic mount which rests in the corner of the glass cell.}
	\label{fig:fig5}
\end{figure}

The glass cell is shown in fig.~\ref{fig:fig5}\,(a). Inside the glass cell a super-polished prism is mounted in a ceramic mount (Macor) which sits in the cell on three legs. The prism is a standard Dove-prism (Melles Griot 01PDE405) and has been super-polished (Coastline Optics) to a surface roughness of $<1~{\rm \AA}$ on the front face. The mount and prism were simply placed in the glass cell and are held in place in the corner of the cell by friction and gravity alone.
\section{Magnetic transport}\label{sec:Magnetictransportoveranobstacle}
In this section we first introduce the basic ideas of magnetic transport and the technical construction of our magnetic transport system. Then in subsection A we describe the loading of the transport trap with ultracold atoms and the principle of their transportation past the obstacle prism to the glass cell. In subsection B we calculate the exact spatial trajectory around the obstacle prism before proceeding in subsection C to describe experiments that we used to characterize and optimize the transport process.

It is well established that low-magnetic-field-seeking atoms can be trapped by a local minimum in a static magnetic field \cite{bergeman_erez_87}, such as the quadrupole potential formed by two coils in anti-Helmholtz configuration. The magnetic moment of the atoms stay aligned with the local field by Larmor precession so long as the local field is large enough, typically greater than 0.2~G. Majorana losses due to spin flips to untrapped states occur close to the field zero at the trap center; this only becomes problematic for very cold ensembles~\cite{Petrich1995}.

If the trapping potential is moved sufficiently slowly, that is (intuitively) with acceleration much less than the native acceleration experienced by the atoms due to the potential, then the atoms may be transported over large distances of many centimeters along with the potential. The trapping potential may be translated either by physically moving the field coils, as was first demonstrated using coils mounted on a motorized translation stage by the group in JILA~\cite{Lewandowski2003}, or alternatively by cross-ramping the currents of a row of static multiple trap coils~\cite{Greiner2001}. The magnetic transport technique has been proven to move cold atoms with negligible heating~\cite{Lewandowski2003,Nakagawa2005,Klempt2008,Jesper2007}.

Our magnetic transport system uses a pair of coils in near-anti-Helmholtz configuration which are displaced using a motorized translation stage (Parker XR404) that runs along a robust rail track of length 80~cm. The transport coils are constructed from Kapton-insulated, square cross-section, hollow copper tubing and have 8$\times$3 turns per coil. The geometry of the transport coils is constrained mainly by the previously minimized height of the MOT chamber. The coils have an inner diameter of 60~mm and and have an inner separation of 87~mm. Water cooling of the coils is necessary to dissipate the heat produced by the large currents we use; up to 400~A to produce an axial gradient of 240~G\,cm$^{-1}$. The current for the transport coils is provided from an Agilent 6690A power supply in constant voltage mode. An external feedback control with a closed-loop servo is used in combination with digital and analog outputs created by an FPGA Labview control system. The feedback loop comprises a Hall effect current sensor~\cite{Honeywell} and a bank of five parallel field effect transistors placed in series with the coils.

The coil movement is controlled via proprietary software which can be programmed to produce any sequence of consecutive movement `profiles'. A single profile consists of three stages: an acceleration stage from zero velocity, a constant velocity stage and a deceleration stage back to zero velocity. Hence a plot of velocity versus time has a trapezoidal shape. The maximum possible acceleration and velocity are 4~m\,s$^{-2}$ and 4~m\,s$^{-1}$ respectively. The positioning at the end of a movement profile is accurate to $\pm10-50$~$\mu$m depending on the deceleration used.

Using a quadrupole trap to transport the atoms along the vacuum system has the advantage that the trap center can be easily displaced by applying a bias field. Thus we are able to bend the path of the magnetic potential around the obstacle prism as the atoms are transported from the MOT to the glass cell. For this purpose we use a pair of `shift' coils (see fig.~\ref{fig:fig4}) which have 3$\times$3 turns per coil. These coils can produce bias fields of up to 250~G, sufficient to shift the trap center by 1~cm vertically at our highest operating transport gradient of 250~G\,cm$^{-1}$. The dimensions of the shift coils were physically constrained both by the size of the enclosed vacuum system and by need for the transport coils to pass over them. The shift coils can be modeled by a pair of equivalent single turn coils with radii of 27(1)~mm and separated by 47(1)~mm.

\subsection{Loading ultracold atoms into the transport trap}
To start an experimental procedure we use standard laser cooling and trapping techniques to collect atoms into a MOT from a background vapor, apply a sub-Doppler cooling phase and then spin-polarize the gas by optical pumping in order to obtain a low-field-seeking, magnetically trappable sample.

Our laser setup consists of two commercial extended cavity diode lasers (Toptica DL 100) and a tapered amplifier (Toptica BoosTA). Both the diode lasers operate on the 780\,nm 5S$_{1/2}$ $\rightarrow$ 5P$_{3/2}$ transition. The first laser generates the light for laser cooling and imaging and is stabilized to the `cycling' transition ($F = 3 \rightarrow F' = 4$ for $^{85}$Rb) using modulation transfer spectroscopy~\cite{MCCarron2008}. The second laser generates the light for repumping and optical pumping and is stabilized to the `repump' transition ($F = 2 \rightarrow F' = 3$ for $^{85}$Rb) using frequency-modulation spectroscopy~\cite{Gehrtz1985}. Acousto-optical modulators (AOMs) centered at 110~MHz are used for intensity  and frequency control of the trapping beams. Optical pumping light resonant with the $F = 2 \rightarrow$ $F' = 2$ transition is generated via an AOM. Light from the laser table is transmitted to the experiment via a total of five polarization maintaining fibers. The $^{85}$Rb MOT uses a total of 90~mW for trapping and cooling and 10~mW for repumping in beams of a final 1/e$^2$ diameter of 30~mm.

In the MOT we typically load up to 1$\times$10$^9$ $^{85}$Rb atoms at a gradient of $10\,\mbox{G$\,$cm$^{-1}$}$. In a 20~ms compressed MOT stage we simultaneously relax the MOT to a gradient of $5\,\mbox{G$\,$cm$^{-1}$}$, change the detuning to $-30$~MHz~\cite{Cornell1994} and decrease the repump intensity to put more atoms in the dark $F = 2$ state and hence reduce radiative repulsion. This stage is followed by a molasses stage of 15~ms with a detuning of $-50$~MHz. In the next step the atoms are optically pumped to the magnetically trappable $F = 2, m_F = -2$ state and the quadrupole field is switched on at $45\,\mbox{G$\,$cm$^{-1}$}$ axially. The magnetic trap then ramps up adiabatically to $180\,\mbox{G$\,$cm$^{-1}$}$ in 500~ms.

The transport coils are then moved over a distance of 51~cm to the science chamber in about 2.5 seconds. In the science chamber the atoms are transferred into a static quadrupole trap produced by the local quadrupole coils, by overlapping the centers of both traps and simultaneously cross-ramping the current of the two traps adiabatically in 500~ms. At the end of this transport process, standard absorption imaging techniques are used to probe the temperature and density of the atomic cloud. The probe pulse is 10~$\mu$s in duration, resonant with the $F = 3 \rightarrow$ $F' = 4$ transition. There are sets of shim coils around both the MOT chamber and the science cell to compensate for any stray magnetic fields present in the laboratory and to produce small bias fields for optical pumping where needed. These coils are made out of 1~mm insulated copper wire, and can typically produce fields of up to 6~G.

The delivery of a high atom number cloud at a low temperature from the MOT chamber into the science cell is an essential prerequisite to cooling atoms further down towards quantum degeneracy. Thus it is important to minimize unwanted heating of the atoms during transportation. One possible source of such heating is poor spatial alignment of the transport magnetic trap with the cloud of cold atoms after the molasses phase. If the center of mass of the atoms is initially displaced from the trap zero at the time of loading, `sloshing' motion occurs and leads to both loss and heating. Loss occurs when an off-axis atomic cloud is clipped passing through the 6~mm tube and/or the 5~mm aperture in the differential pumping stage. Heating occurs as the center of mass motion is converted into random ensemble kinetic energy. To minimize this problem we measured the final cloud temperature after transport as a function of the initial transport trap position in each cartesian direction. This was done by physically moving the transport coil system whilst monitoring its position with a micrometer gauge. Clear minima in the temperature were found for each direction, giving the optimum loading position.  We found a difference of up to 25\% increase in temperature if the transport trap is misaligned by 4~mm along the direction of transport. Another potential source of sloshing/heating occurs when the atoms are moved over the obstacle prism and this is examined in detail in the next two subsections.
\subsection{Taking atoms past the obstacle prism}\label{sec:PassingObstacle}
\begin{figure}
	\centering
		\includegraphics[width=0.40\textwidth]{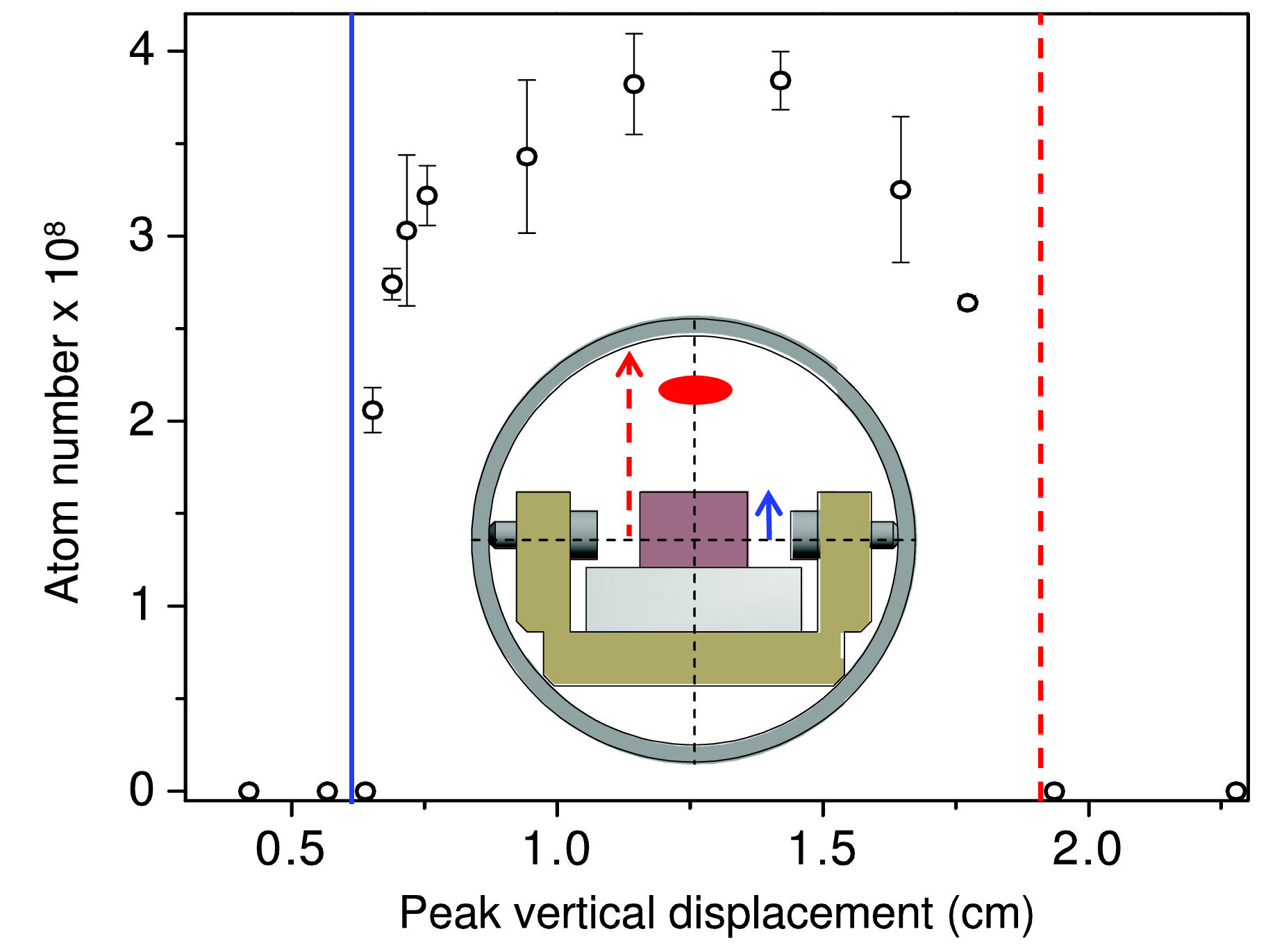}
	\caption{Atom number as a function of the peak vertical displacement of the trap transporting the atoms. On the left hand side (blue solid line) of the graph atoms are lost as they hit the prism, on the right hand side atoms are lost due to hitting the walls of the vacuum tube (red dashed line). Inset shows a drawing of the obstacle prism and mount looking along the transport path. The red ellipse shows the vertical displacement of the atom cloud as it passes over the obstacle.}
	\label{fig:fig6}
\end{figure}
When transporting atoms from the MOT to the glass cell a bias field produced by the shift coils around the obstacle deflects the atoms along a curved trajectory inside the vacuum chamber but passing over the obstacle, see fig.~\ref{fig:fig6}. The number of atoms transported over the obstacle is shown as a function of the peak vertical displacement of the transport trap field zero. On the left hand side of the blue line atoms are lost due to not passing over the obstacle. On the right hand side of red dashed line atoms are lost due to hitting the vacuum tube. When transporting atoms in a quadrupole trap with an axial field gradient of  $180\,\mbox{G$\,$cm$^{-1}$}$, we apply a bias field of $\sim$216~G using the shift coils resulting in a 12~mm peak vertical displacement.

It is useful to know the exact trajectory of the trap to ensure that it is smooth and does not pass too near to the vacuum system walls or mounting components. Hence we calculate the trap displacement in the z-direction (vertical) and x-direction (horizontal in the direction of transportation) as a function of the distance between the moving transport coil and static shift coil centers. 
\begin{figure}
	\centering
		\includegraphics[width=0.40\textwidth]{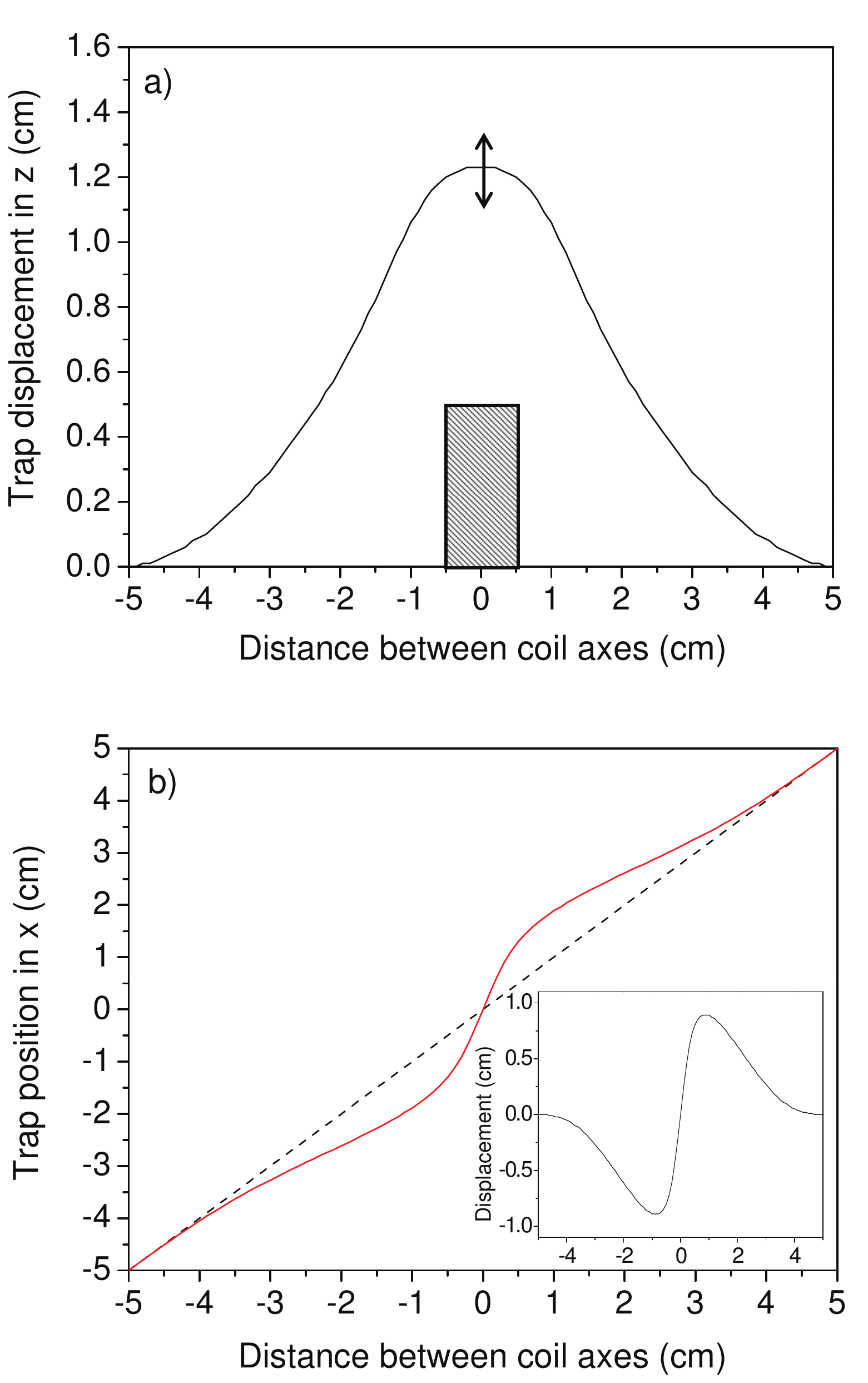}
	\caption{Displacement of trap center during the transport over the obstacle. a) Vertical (z) trap displacement as a function of distance between the coil centers. The shaded rectangle shows the size of the obstacle and the arrow show the typical rms vertical diameter of a 200~$\mu$K cloud. b) Horizontal (x) position of the trap center (red solid line) compared to the position of the transport coil center (black dashed line). The inset shows the trap displacement from the transport coil center as a function of distance between the coil axes.}
	\label{fig:fig7}
\end{figure}
The magnetic field generated by each coil is calculated by numerically integrating the Biot-Savart law and the contributions from each coil are summed. The calculations are performed for various distances between the coil axes. As shown in fig.~\ref{fig:fig7}\,(a) the trap gets smoothly displaced in the z direction whilst the transport coils pass around the shift coils, in agreement with the data in fig.~\ref{fig:fig6}. For the x-direction, shown in fig.~\ref{fig:fig7}\,(b), the trap position [solid red line] differs from the coil position [dashed black line]. When the transport coils are approaching the shift coils the trap center lags behind the position of the transport coils by as much as 1 cm. The inset of fig.~\ref{fig:fig7}\,(b) shows the trap x-displacement from the transport coil center as a function of the distance between the transport and shift coil axes. The trap behaves as if it is `repelled' by the obstacle in both the z- and x- directions and thus the atoms follow a smooth, clear path over the obstacle.
\subsection{Optimizing the transport process}
In this section we describe experiments carried out to analyze and optimize the transport process. Having optimized the loading of the transport trap, we studied the effect of varying the transport velocity profiles on the final number and temperature of the transported atoms. The two variable parameters are the acceleration/deceleration and the constant velocity of the trapezoidal profile.

The first experiment carried out was to vary the acceleration whilst keeping the intermediate velocity constant. The velocity was kept at 0.26~m\,s$^{-1}$ and the acceleration and deceleration were changed pairwise. We found that the atom number is not dependent on the acceleration up to a value of 2.75~m\,s$^{-2}$ which is about 1/15th of the trap's native radial acceleration. Above that value the atom number started to decrease by as much as a factor of two with an associated increase in final temperature from 200 to 240~$\mu$K. Hence in practice we restrict accelerations to the lower, safe range around 1~m\,s$^{-2}$ which allows the entire transport process to be completed in just 2.25 seconds with $\sim 70\%$ of the atoms loaded being transferred. As the bulk of the transport time is spent in the constant velocity stage, there is little to be gained by using higher acceleration. We believe that the loss and heating seen at higher accelerations is caused by excitation of small sloshing motions that lead to clipping of the atom cloud in the narrow tube at the start of the differential pumping stage.

\begin{figure}
	\centering
		\includegraphics[width=0.40\textwidth]{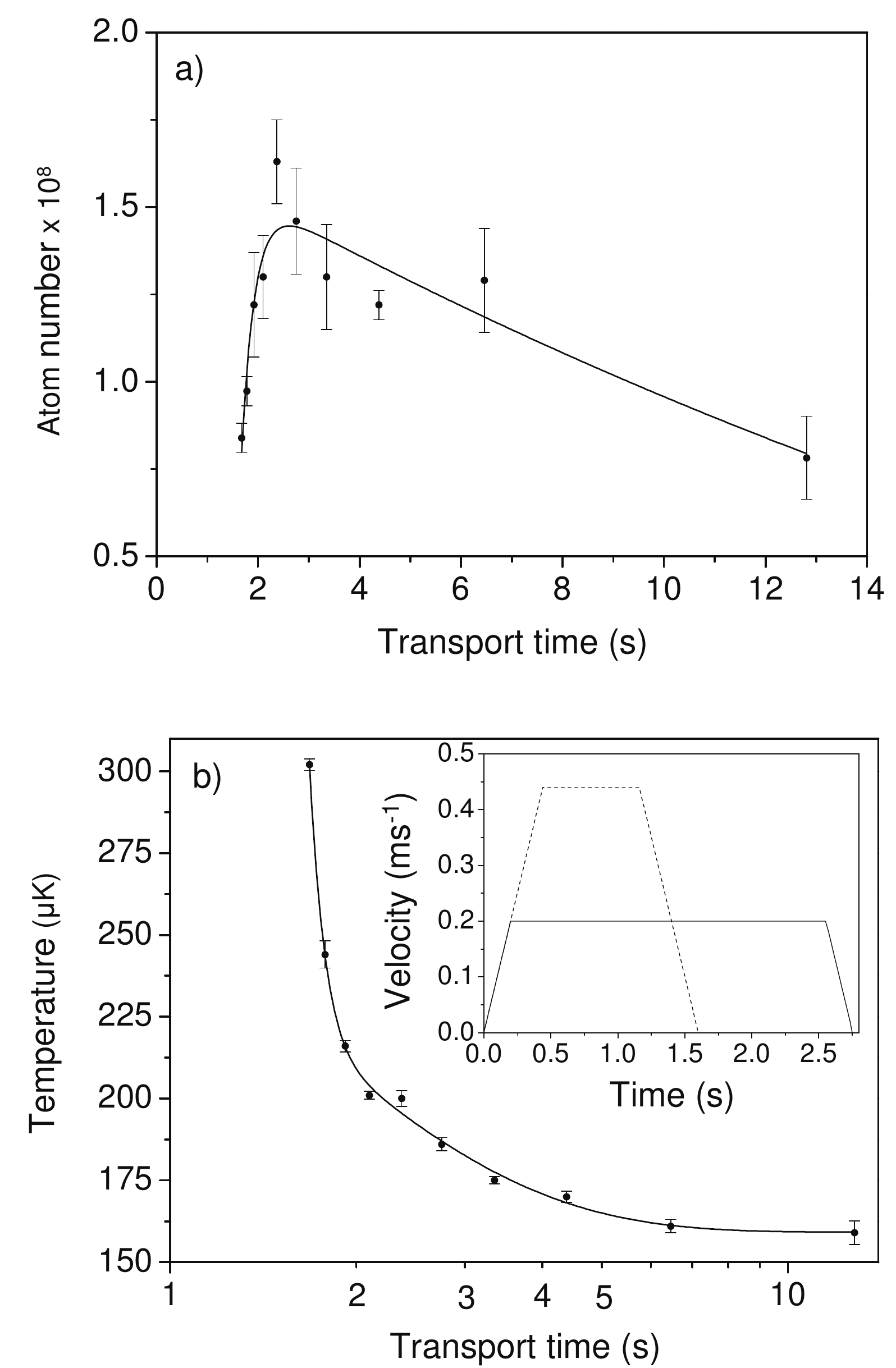}
	\caption{Transported atom number (a) and temperature (b) as a function of the transport time. For this experiment the acceleration and deceleration were kept constant at 1.0~m\,s$^{-2}$ and the velocity of the trapezoidal profile was varied [see inset of (b)]. The solid lines through the data are to guide the eye only.}
	\label{fig:fig8}
\end{figure}

In a second experiment the acceleration and deceleration were kept constant at 1.0~m\,s$^{-2}$ and the velocity was varied. In fig.~\ref{fig:fig8}\,(a) and (b) the resulting atom number and temperature are shown as a function of the transport time. The velocity profile was always trapezoidal [see inset fig.~\ref{fig:fig8}\,(b)]. For very short transport times $<2$~s, i.e. high velocities, we observed losses and heating of the atomic cloud. Increasing the transport time to 2.2~s (i.e. reducing the speed to 0.26~m\,s$^{-1}$) reduced the loss and heating and gave the optimum transfer. Increasing the transport time further led again to losses. This is because the atoms spend too long in the high pressure MOT chamber and are subject to collisional losses from the background gas as was shown earlier in fig.~\ref{fig:fig3}. In fig.~\ref{fig:fig8}\,(b) the temperature is lower for longer transport times, we believe this is due to hot atoms of the cloud being lost through collisions with the tube walls: an unintended form of evaporative cooling.

\begin{figure}
	\centering
		\includegraphics[width=0.40\textwidth]{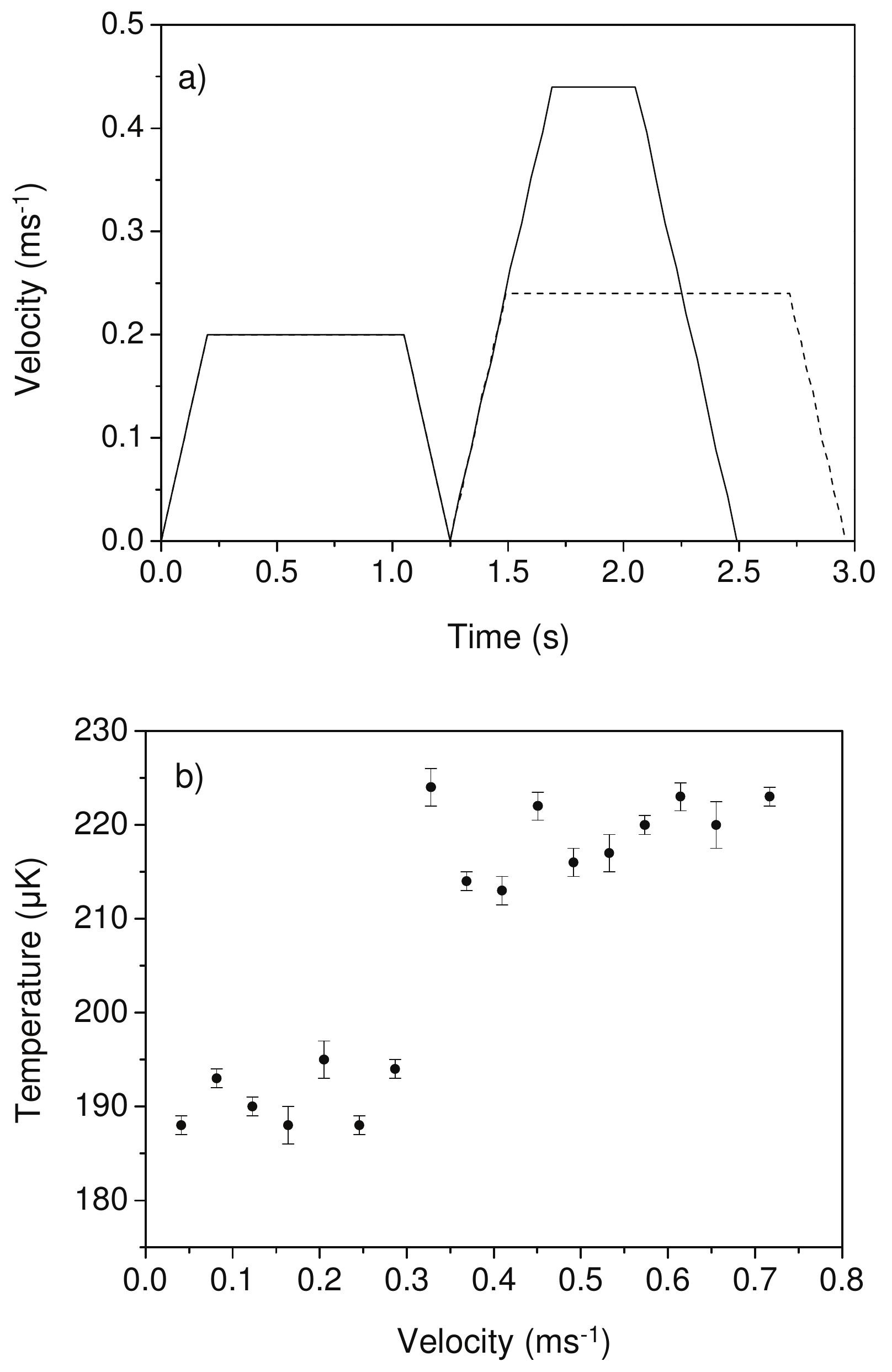}
	\caption{Measurement of the temperature as a function of transport velocity over the obstacle. a) Example of a two-part transport profile: the transport before the obstacle is kept the same, and only the velocity over the obstacle in the second part of the profile is changed (as indicated).  b) For transport velocities above 0.3~m\,s$^{-1}$ there is a sudden increase in the temperature of the cloud.}
	\label{fig:fig9}
\end{figure}

\begin{figure}
	\centering
		\includegraphics[width=0.40\textwidth]{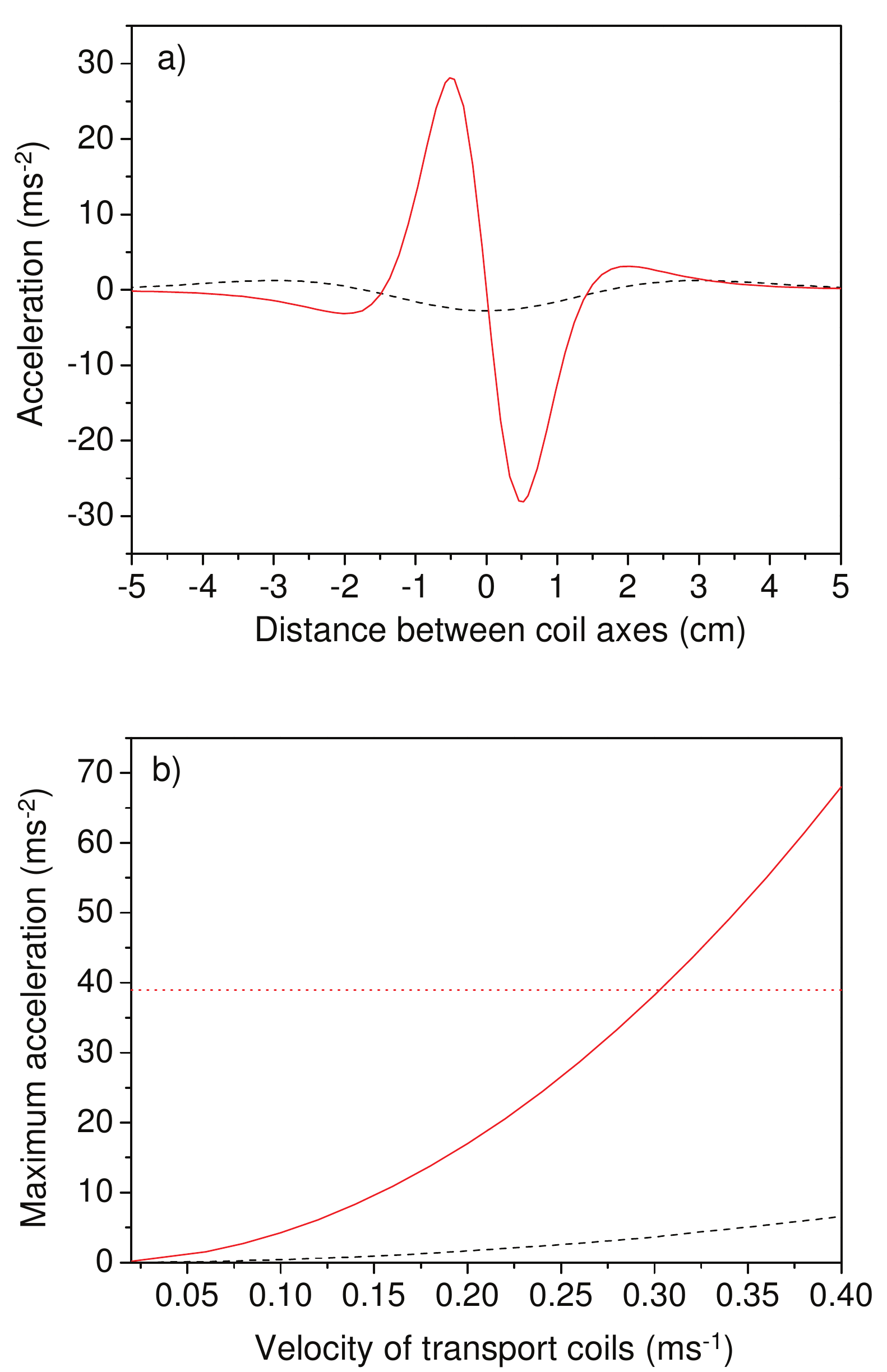}
	\caption{Accelerations during the transport over the obstacle. a) Acceleration in z (black dashed line) and x (red solid line) of the trap center as a function of distance between the coil axes calculated for a transport velocity of 0.26~m\,s$^{-1}$. b) Maximum acceleration in z (black dashed line) and x (red solid line) for various transport velocities. The dotted red line marks the native radial acceleration of the atoms in a quadrupole trap with a 180~G\,cm$^{-1}$ vertical field gradient.}
	\label{fig:fig10}
\end{figure}

We hypothesized that the loss and heating at short transport times were caused by going over the obstacle too fast and in order to test this we employed a two part velocity profile. This allowed us to separate effects due to transport over the obstacle from those due to transport out of the MOT chamber and along the differential pumping stage. A typical two part profile is shown in fig.~\ref{fig:fig9}\,(a): in the first part the transport occurs with a constant speed until the atoms have passed the second aperture where the transport comes to rest momentarily. (A limitation of the software used to control the motorized translation stage restricts the motion to a sequence of trapezoidal profiles.) This part of the transport was fixed and we only varied the final velocity of the second part (indicated by the red line), which is the velocity over the obstacle. The results are shown in fig~\ref{fig:fig9}\,(b), where we see that for velocities smaller than 0.3~m\,s$^{-1}$ the temperature is consistently just below 200~$\mu$K. However for velocities higher than 0.3~m\,s$^{-1}$ we see a stepped increase in the temperature, confirming that the transport coils are going `too fast' over the obstacle.
We now compare this result with our earlier calculation in section~\ref{sec:PassingObstacle}. Knowing the trap displacement in the x and z direction and assuming that the transport coils move past the obstacle at a constant speed we can calculate the acceleration which the atoms will experience as the trap passes the obstacle. Contrary to intuition, the accelerations along the transport (x) direction are nearly an order of magnitude larger than accelerations due to the vertical (z) displacement [see fig.~\ref{fig:fig10}\,(a)] for a transport velocity of 0.26~m\,s$^{-1}$. In fig.~\ref{fig:fig10}\,(b) we plot the peak accelerations in the x and z directions for different transport velocities. The black dashed line shows the maximum acceleration in the z direction which is small compared to the native axial acceleration of the atoms due the linear trap potential; this is 78~m\,s$^{-1}$ for 180~G\,cm$^{-1}$. In contrast, we find that for transport velocities larger than 0.3~m\,s$^{-1}$, the acceleration along the x direction (solid red line) is greater than the native radial acceleration of 39~m\,s$^{-1}$ (red dotted line). This is consistent with the experimental observations and provides a satisfactory explanation for the sudden onset of loss and heating at higher velocities due to excitation of large sloshing motion. Moreover our measurements and modeling demonstrate that there is a large parameter range where the atoms can be safely transported over the obstacle with negligible loss and heating.
\section{Loading of a hybrid trap and production of $^{87}$Rb condensates}\label{sec:HybridTrap}
In order to demonstrate the overall performance of the apparatus we produce a Bose-Einstein condensate of $^{87}$Rb in a hybrid trap, this being a combination of optical radial and magnetic axial confinement that has been successfully employed in former work~\cite{Davis95,Hoang05,Hung08,Porto2009,Jenkin2010}. Initially the atoms are loaded into a purely magnetic trap formed by the science quadrupole coils, where they are cooled by RF evaporation before transfer to the hybrid trap. The dipole laser beam is brought into the cell through the back of a super-polished dove prism which is located in the glass cell. The atoms are confined radially by this single beam dipole trap and axial confinement along the beam is produced by the magnetic potential of the quadrupole coils (see fig.~\ref{fig:fig5}). 

The experimental procedure for $^{87}$Rb up to the RF evaporation is similar to that described in section~\ref{sec:Magnetictransportoveranobstacle} for the transport experiments with $^{85}$Rb, however we briefly summarize the process as parameters for $^{87}$Rb are slightly different. 
The MOT is loaded for 25~s and typically fills to 6(1)$\times 10^8$ atoms. The atoms are optically pumped into the $F = 1, m_F = -1$ stage and loaded into the transport magnetic trap at an axial field gradient of 60~G\,cm$^{-1}$. This is larger than the gradient used for $^{85}$Rb by 4/3, the ratio of the two isotopes' magnetic moments, and ensures the same degree of confinement in the transport trap. The gradient is adiabatically ramped up to 225~G\,cm$^{-1}$ before initiating the transport process. The magnetic field from the shift coils is adjusted to deliver the atoms over the obstacle. After the transport we have 2.3(1)$\times 10^8$ atoms at a temperature of 240(1)~$\mu$K and with a PSD of $10^{-7}$ in the science chamber. The lifetime at this stage is 190(10)~s. The atoms are then cooled by forced RF evaporation with three RF ramps between the frequencies of 40~MHz and 5~MHz in 30~s to obtain 2.5$\times 10^7$ atoms at 29~$\mu$K with a PSD of $6\times10^{-5}$. During the RF evaporation the dipole trap beam is turned on, it does not perturb the RF evaporation. At the end of the evaporation the lifetime is limited to 23(4)~s owing to Majorana spin-flips at the field zero. To proceed we load the hybrid trap, which has an initial depth of 70~$\mu$K. The transfer to the hybrid trap is made by ramping down the magnetic field gradient to 29~G\,cm$^{-1}$ (just below that needed to support the atoms against gravity) whilst keeping the RF frequency at 5~MHz. The dipole trap loads through elastic collisions in a similar way to a dimple trap, described in former work~\cite{Porto2009,Kraemer04,Stamper-Kurn98,KleineBuning2010}.
\begin{figure}
	\centering
		\includegraphics[width=0.48\textwidth]{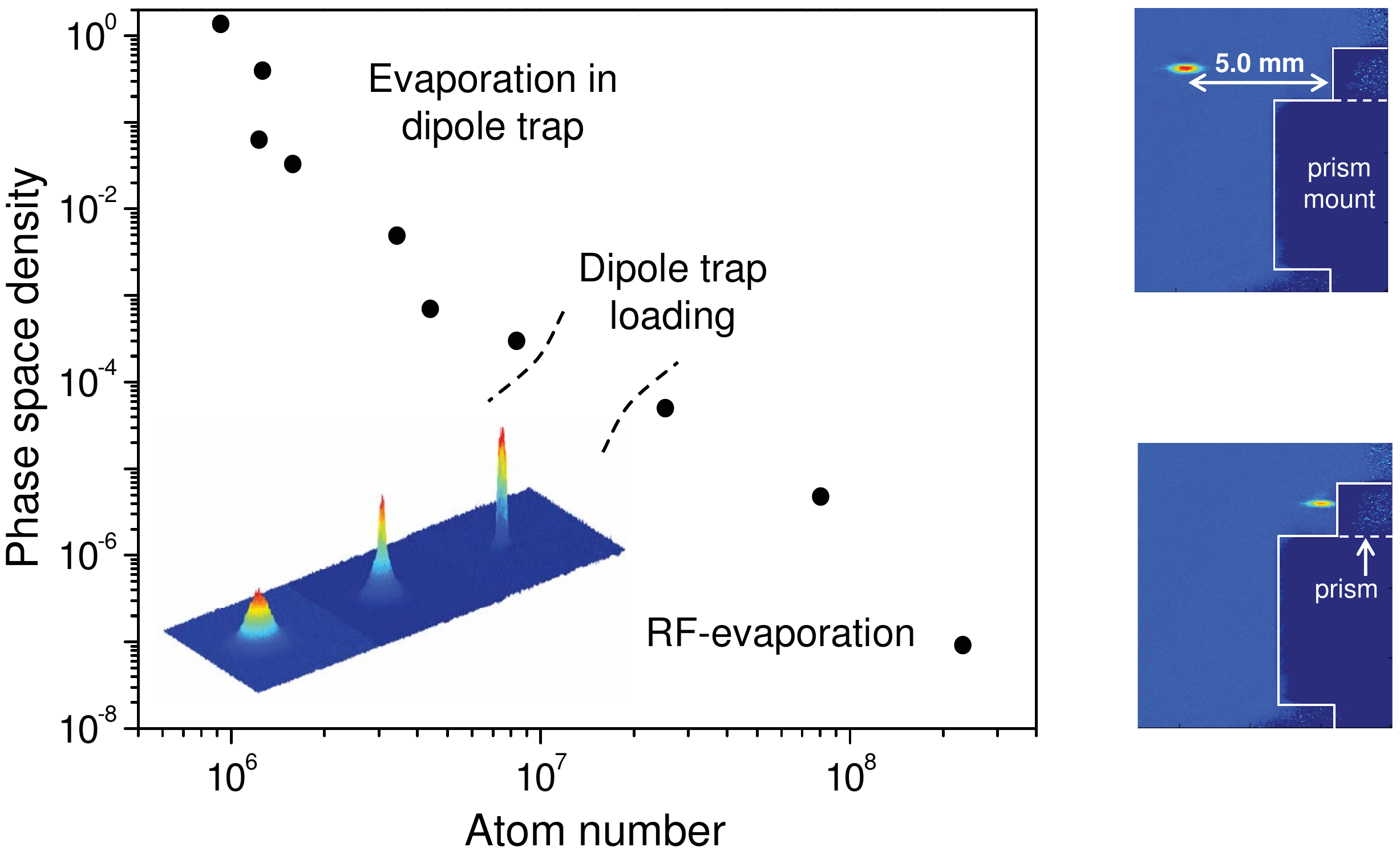}
	\caption{The trajectory to BEC in a hybrid trap starting from N $\sim2\times10^{8}$, PSD $\sim10^{-7}$ in the bottom right and progressing to N $\sim10^{6}$, PSD $>1$ in the top left. The inset shows shows false color images of the transition to BEC from a thermal cloud (L) to a bimodal distribution to a pure condensate (R). Right hand side: Absorption images of cold atoms which are brought up to the surface.}
	\label{fig:fig11}
\end{figure}
The evaporation to quantum degeneracy is achieved by reducing the trap depth linearly in time. In fig.~\ref{fig:fig11} the phase space density is plotted as a function of atom number for the various stages of evaporation. Experimentally we reduce the beam powers from initial 2.5~W to 120~mW at the final stage. The trap depth in the final, shallow trap is 3~$\mu$K and the frequencies are $\omega_z = 17$~Hz, $\omega_x=\omega_y = 165$~Hz. At the end of the intensity ramps we observed the onset of BEC at a temperature of $\sim$200~nK.

Another attractive feature of the hybrid trap is the ease with which trapped atoms may be moved along the waveguide provided by the dipole beam through use of a horizontal magnetic bias field; this shifts the center of the axial magnetic confinement. The two absorption images on the right of fig.~\ref{fig:fig11} illustrate the use of this technique to move the atoms towards and against the prism surface. We have elsewhere demonstrated \cite{Marchant2011} the use of such techniques to measure ultracold atom cloud temperatures and also to evaporatively cool atoms to degeneracy by using the prism surface to selectively remove hot atoms.

\section{Conclusion}\label{sec:Conclusion}
We have described an apparatus designed for the study of ultracold atoms near a room temperature surface. The system employs magnetic transport in order to give excellent access near to the surface. This has necessitated the design of a low profile MOT chamber which allows the coils generating the quadrupole trap for transport to be mounted close to the transport axis, whilst keeping good optical access for large diameter laser beams. Additionally we have included an obstacle which, in conjunction with the differential pumping elements, blocks the line of sight between the MOT chamber and the surface. We have described how atoms can be easily transported over such an obstacle with negligible loss and have presented details of the transport optimization. Finally we demonstrate the production of $^{87}$Rb condensates and their transport up to to the surface. Future work will focus on the production of $^{85}$Rb condensates close to the surface where the presence of a broad Feshbach resonance permits the intriguing prospect for the study of bright matter wave solitons reflecting from the surface~\cite{Cornish2009}. 
\section{Acknowledgments}
We acknowledge support from the UK Engineering and Physical Sciences Research Council (grants EP/F002068/1 and EP/G026602/1) and the European Science Foundation within the EUROCORES Programme EuroQUASAR. SLC acknowledges the support of the Royal Society.


\end{document}